\documentclass[12pt,psf,epsf]{JHEP3}

\usepackage{graphicx}
\usepackage{graphics}
\usepackage{epsfig,bm}
\usepackage[small]{caption}
\usepackage{amsfonts,amssymb,amsmath,latexsym}

\newcommand{\be}{\begin{equation}}
\newcommand{\ee}{\end{equation}}
\newcommand{\bea}{\begin{eqnarray}}
\newcommand{\eea}{\end{eqnarray}}

\title{Holography and non-local operators in the BTZ black hole with non-zero angular momentum.
}

\author{D. S. Ageev \\
\it Steklov Mathematical Institute, Russian Academy of Sciences,\\
Gubkina str. 8, 119991, Moscow, Russia \\
ageev@mi.ras.ru}

\author{I.Ya. Aref'eva \\
\it Steklov Mathematical Institute, Russian Academy of Sciences,\\
Gubkina str. 8, 119991, Moscow, Russia \\
arefeva@mi.ras.ru}

\maketitle

\abstract{
We study quark-antiquark potential using the AdS/CFT correspondence in the BTZ black hole with non-zero angular momentum.
Using explicit form of string configurations relevant to a calculation of the potential we find that the potential exhibits
  different dependencies on angular momentum values in the Euclidean and the Lorentzian signatures of the BTZ.}


\begin{document}
\maketitle
\tableofcontents
\section{Introduction}

It's well known that the study of the behaviour and properties of quark-gluon plasma (QGP), obtained
in heavy ion collisions  at RHIC and LHC, cannot be performed by perturbative QCD. A holographic
approach to this purpose is widely  appreciated  and developed, see reviews \cite{CasalderreySolana:2011us,Aref'eva:2012kr,DeWolfe:2013cua}. Although the exact AdS dual to QCD is not
known, the gauge-string duality is one of the
fruitful  ways to study QGP \cite{CasalderreySolana:2011us}, since  the finite-temperature field theory that arises in the AdS/CFT correspondence shares
many properties of  QGP \cite{CasalderreySolana:2011us,Aref'eva:2012kr,DeWolfe:2013cua}.

The spatial anisotropy is an important characteristic of QGP, obtained from  heavy ion collisions experiments, and there are several proposals
to describe it through the direct application of the AdS/CFT correspondence.
One of the ways to take anisotropy into account is to deal with Kerr black holes in AdS background \cite{NataAtmaja:2010hd,
McInnes:2012gw,Aref'eva:2013wma}
(about AdS/CFT for Kerr-AdS see, for example \cite{Hawking:1998kw,Hawking:1999dp}).
 Another way  is to deal directly with an anisotropic background
 \cite{Giataganas:2013lga}.
  In \cite{NataAtmaja:2010hd} leading corrections to the quark drag-force  caused by the Kerr-rotating parameter $a$
   have been calculated. Meanwhile, in
  \cite{Aref'eva:2013wma} it has been shown that the time of thermalization
  does not depend on $a$ in 3-dimensional case. In this paper we are going to study the dependence of
  loops correlators on this rotating parameter $a$.

Polyakov  and Wilson loops correlators provide natural and
simple probes in studies of temperature gauge theories and QGP.
 A lot of papers  considered the AdS/CFT duality to elaborate on these non-local operators and use them to investigate the
 behaviour of the quark-antiquark pairs potentials in the different cases \cite{Maldacena:1998im,
 Brandhuber:1998bs,Rey:1998bq,
 Liu:2006he,Jugeau:2012bf}.
The temporal Wilson loop at the zero-temperature has been calculated
using the AdS/CFT in the pioneering paper \cite{Maldacena:1998im} and the
non-zero temperature Polyakov loop correlators  have been calculated in  \cite{Brandhuber:1998bs,Rey:1998bq}, see also refs. in
\cite{CasalderreySolana:2011us,DeWolfe:2013cua}.
The discussion of the adjoint interquark potential in holographic context can be
 found in \cite{Bak:2007fk, Albacete:2008dz, Grigoryan:2011cn}.

In the present paper we study the holographic Wilson and
Polyakov loops in the BTZ-black hole with non-zero angular momentum and  extract interquark interaction potentials from these quantities.
In accordance with the holographic prescription
we consider the Lorentzian version of the BTZ for the Wilson loop and the Euclidean one for the Polyakov lines and  obtain corresponding  potentials.

The potential of the interquark interaction is defined by the action of
U-shaped string
configuration ending at quarks and hanging in the corresponding space up to a maximal coordinate $z_m$.
We find that the U-shaped string action rewritten in terms of this maximal coordinate $z_m$ doesn't contain
 explicit dependence on rotation parameter $a$ for both signatures. Note, that we do not consider the possible topological effects related to the structure of the BTZ.
The whole dependence on $a$ is accumulated in dependence of $z_m$ on $a$.
For non-rotating black holes  dependencies of the maximal $z_m$ on
the  interquark distance $d$
are also the same for both signatures,
so the singlet potentials are coincide.
For a rotating black hole this coincidence is not valid anymore.
We find that
the potentials exhibit a dependence on angular momentum and this dependence increases for high values of $a$.

The paper is organised as follows.
In section 2 we remind the BTZ black hole geometry in the Lorentzian and Euclidean
versions and the holographic prescription for the Wilson and Polyakov  loops.
In section 3 and 4 we derive the formulae for the maximal hanging distance for both signature cases.
In section 3 we show, that there are restrictions on $z_m$ defining a "living space" for a string configuration connecting two quarks.
In section 5 we calculate the potential and the action of the static quark-antiquark pair.
In section 6 we use  complex-valued single string configurations to perform renormalization for the Lorentzian signature case and compare this renormalization with renormalization in the Euclidean case.
In Appendix   we discuss  approximations for integrals used in the main text.

\section{The set-up}
\subsection{The BTZ background.}

We start with the metric of the BTZ black hole (in the Lorentzian version)
in the coordinates $(r,t,\phi)$

\begin{equation}
ds^{2}=-(-M+\frac{r^{2}}{l^{2}})dt^{2}+\frac{dr^{2}}{-M+\frac{r^{2}}{l^{2}}+\frac{a^{2}}{r^{2}}}-2a dtd\phi+r^{2}d\phi^{2},
\end{equation}
where $-\infty<t<\infty$,\, $0<\phi<2\pi$,\, $r>0$,
$M$ is the mass of the BTZ black hole, $a$ is its angular momentum and $l$ is the scale parameter.
 We set $l=1$ until the end of the paper.

One can rewrite this metric using the change of variable $r=\frac{1}{z}$

\begin{equation}\label{BTZ_nE_z}
ds^{2}=-(-M+\frac{1}{z^{2}})dt^{2}+\frac{1}{z^{4}}\frac{dz^{2}}{-M+\frac{1}{z^{2}}+a^{2}z^{2}}-2a dtd\phi+\frac{1}{z^{2}}d\phi^{2}.
\end{equation}

The Euclidean version of  metric \eqref{BTZ_nE_z} is

\begin{equation}\label{BTZ_E_z}
ds_E^{2}=(-M+\frac{1}{z^{2}})dt^{2}+\frac{1}{z^{4}}\frac{dz^{2}}{-M+\frac{1}{z^{2}}
-a^{2}z^{2}}+2adtd\phi+\frac{1}{z^{2}}d\phi^{2}.
\end{equation}

Horizons of metric \eqref{BTZ_nE_z} are obtained as solutions of the following equation

\begin{equation}
-M+\frac{1}{z^{2}}+a^{2}z^{2}=0.
\end{equation}
We get horizons for the Lorentzian version of metric \eqref{BTZ_nE_z} in the $z$-coordinate as

\begin{equation}\label{horizon1}
z_{+}={\displaystyle \frac{1}{2}{\frac{\sqrt{2}\sqrt{M+\sqrt{{M}^{2}-4\,{a}^{2}}}}{a}}},
\end{equation}

\begin{equation}\label{horizon2}
z_{-}={\displaystyle \frac{1}{2}{\frac{\sqrt{2}\sqrt{M-\sqrt{{M}^{2}-4\,{a}^{2}}}}{a}}},
\end{equation}
and for the Euclidean version of metric \eqref{BTZ_E_z} in the $z$-coordinate as

\begin{equation}\label{horizon1e}
z^{(E)}_C= \frac{1}{2}\frac{\sqrt{2M+2\sqrt{M^2+4\,a^2}}}{ia},
\end{equation}

\begin{equation}\label{horizon2e}
z^{(E)}= \frac{1}{2}\frac{\sqrt{2\sqrt{M^2+4\,a^2}-2M}}{a}.
\end{equation}
In the limiting case $a\rightarrow0$, $z_{+}\rightarrow\infty$ and
$z_{-}\rightarrow\frac{1}{\sqrt{M}}$. Note, that $z^{(E)}_C$ is always complex and $z^{(E)}$\,is always real.

\subsection{Holography and the non-local operators.}

\begin{figure}[h!]
    \centering
     \includegraphics[width=10cm]{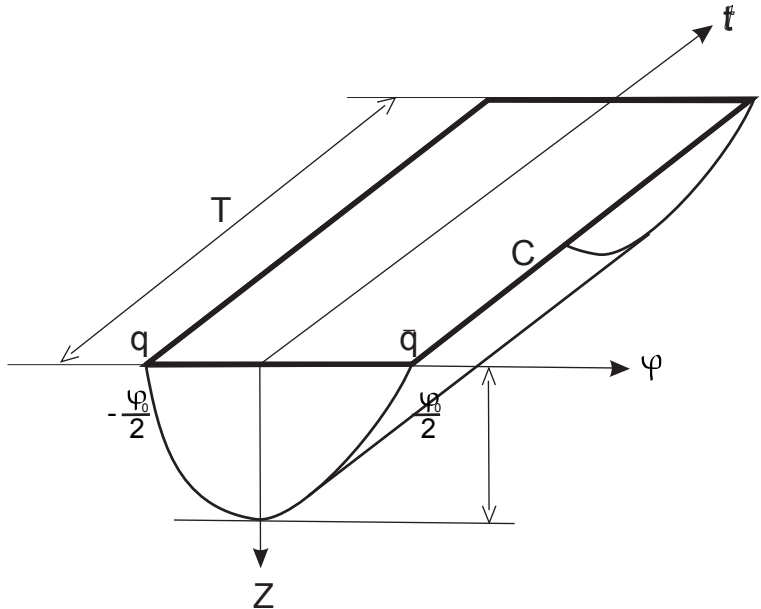}$\,\,\,\,$ $\,\,\,\,$
     \caption{The Wilson loop $C$ corresponding to the static quark-antiquark pair $q$-$\overline{q}$ and the string stretched on $C$ in the BTZ-background. }\label{strstrs}
\end{figure}

According to Maldacena's proposal one calculates the average of the Wilson loop operator $<W(C)>$ using the following formula
\begin{equation}
<W(C)>=e^{-S_{ren}},
\end{equation}
where $C$ is a contour on the boundary of BTZ, $W(C)$ is the Wilson loop operator
\begin{equation}\label{wilson}
W(C)=\frac{1}{N_c} {\mbox {Tr}}({\mbox P}\exp(ig\int\,A_{\mu}dx^{\mu})).
\end{equation}
and  $S_{ren}$ is the renormalized action of the string ending on the contour $S$
and hanging in the BTZ space-time,  see Fig. \ref{strstrs}.

The Euclidean version of the BTZ black hole provides us the  connected correlator of two Polyakov loops
\be<L^{\dag}(-\phi_{0}/2)L(\phi_{0}/2)>=e^{-S_{ren}},
\ee
where $L(r)$ is given by
\begin{equation}
L(r)=\frac{1}{N_c}{\mbox {Tr}}({\mbox P}\exp(ig\intop_{0}^{\beta}A_{4}(r,\tau)d\tau)).
\end{equation}

\subsection{The Nambu-Goto action of the string in the BTZ background.}

The Nambu-Goto action is
\begin{equation}\label{NG1}
S=-\intop d\sigma d\tau\sqrt{-\det(h_{\alpha\beta})},
\end{equation}
where $h_{\alpha\beta}$ is the induced metric
\begin{equation}
\label{ind-lor}
h_{\alpha\beta}=g_{MN}\partial_{\alpha}X^{M}\partial_{\beta}X^{N},
\end{equation}
$g_{MN}$ is the background  metric and $X^{M}(\sigma,\tau)$
is a worldsheet.

The Euclidean version of the action is
\begin{equation}\label{NG2}
S_E=\intop d\sigma d\tau\sqrt{\det(h^{(E)}_{\alpha\beta})},
\end{equation}
where $h^{(E)}_{\alpha\beta}$ is the Euclidean version of the induced metric

\begin{equation}
\label{ind-euc}
h^{(E)}_{\alpha\beta}=g^{(E)}_{MN}\partial_{\alpha}X^{M}\partial_{\beta}X^{N},
\end{equation}
and $g^{(E)}_{MN}$ is the Euclidean version of metric $g_{MN}$.

We choose $\sigma=\phi$, $\tau=t$ and $X^{M}(\sigma,\tau)$ to be
\begin{equation}\label{ansatz}
X^{M}(t,\phi)=\left(\begin{array}{c}
t\\
z(\phi)\\
\phi
\end{array}\right).
\end{equation}
Using \eqref{ansatz} we get the induced metric (\ref{ind-lor}) for metric \eqref{BTZ_nE_z}

\begin{equation}\label{ind_l}
d\Sigma^{2}=-(-M+\frac{1}{z^{2}})d\tau^{2}+(\frac{1}{z^{4}}
\frac{z^{\prime\,2}}{-M+\frac{1}{z^{2}}+a^{2}z^{2}}+\frac{1}{z^{2}})d\sigma^{2}-2ad\tau d\sigma
\end{equation}
and the induced metric (\ref{ind-euc}) for metric \eqref{BTZ_E_z} is

\begin{equation}\label{ind_E}
d\Sigma_E^{2}=(-M+\frac{1}{z^{2}})d\tau^{2}+(\frac{1}{z^{4}}\frac{z^{\prime\,2}}{-M+\frac{1}{z^{2}}-a^{2}z^{2}}+\frac{1}{z^{2}})d\sigma^{2}+2ad\tau d\sigma.
\end{equation}

One can calculate the value of $\sqrt{-\det(h_{\alpha\beta})}$ from expression \eqref{NG1}, where $h_{\alpha\beta}$ is metric \eqref{ind_l}
\begin{equation}\label{det1}
\sqrt{-\det(h_{\alpha\beta})}=\sqrt{{\displaystyle {\displaystyle {\displaystyle {\frac{-{z}^{8}{a}^{4}+2\, M{z}^{6}{a}^{2}+\left(-{M}^{2}-2\,{a}^{2}\right){z}^{4}+2\, M{z}^{2}-1+\left(M{z}^{2}-1\right)z^{\prime\,2}}{{z}^{4}\left(-{a}^{2}{z}^{4}+M{z}^{2}-1\right)}}}}}}.
\end{equation}

We can consider $\sqrt{-\det(h_{\alpha\beta})}$ as a Lagrangian of an one-dimensional system

\begin{equation}\label{action_int}
S=-\intop_{0}^{T}dt\intop_{-\phi_{0}/2}^{\phi_{0}/2}d\phi\sqrt{-\det(h_{\alpha\beta})}
=-T\intop_{-\phi_{0}/2}^{\phi_{0}/2}d\phi\sqrt{-\det(h_{\alpha\beta})}.
\end{equation}

Since Lagrangian \eqref{det1} doesn't contain an explicit dependence on $\phi$, it admits the integral of motion

\begin{equation}\label{P_int}
{\cal {P}}=\frac{1}{z^2}\sqrt{\frac{\left({a}^{2}{z}^{4}-M{z}^{2}+1\right)^3}{\left({a}^{2}{z}^{4}-M{z}^{2}+1\right)^{2}-\left(M{z}^{2}-{1}^{2}\right)(\frac{dz}{d\phi})^{2}}}.
\end{equation}

When $z$ takes its maximal value $z_{m}$ its derivative equals to zero,
so one can write ${\cal P}$  in terms of $z_{m}$

\begin{equation}
{\cal P}={\frac{\sqrt{{a}^{2}{{z_{m}^{2}}}-M{z_{m}^{2}}+1}}{{z_{m}^{2}}}}.
\end{equation}

Therefore, we get the expression for the interquark distance $\phi_{0}$ as the function of $z_m$, $a$ and $M$
\begin{equation}\label{preprefineq1}
\frac{\phi_{0}}{2}=z_{+}z_{-}\intop_{0}^{z_{m}}\sqrt{\frac{\left(M{z}^{2}-1\right)
{\cal F}(a,M,z_{m})}{({z}^{2}-{z_m^2}){\cal G}(M,z,z_{m})}}\,\frac{ z^{2}\, dz}{{\cal F}(a,M,z)},
\end{equation}
where
\bea
\label{F-cal}
{\cal F}(a,M,z)&=&-1+ z^{2}\left(-{a}^{2} z^{2}+M\right)=-a^{2}(z^{2}-z_{+}^{2})(z^{2}-z_{-}^{2}),
\\
\label{G-cal}
{\cal G}(M,z,z_{m})&=&Mz_{m}^{2}z^{2}-z^{2}-z_{m}^{2}.
\eea

 (\ref{preprefineq1}) is the final formula to analyse the string profile behaviour in the Lorentzian case.
To get  an analog of formula \eqref{preprefineq1} in the Euclidean case we make the change in \eqref{preprefineq1} as $a$$\rightarrow$$ia$.

\section{Living space for Nambu-Goto string in BTZ.}

\subsection{ Lorentzian version.}

In this section we derive  restrictions on $z_{m}$ and $z$
 to ensure the positivity of the expression  under the square root in \eqref{preprefineq1} when $z\in(0,z_{m})$ where the integration takes place. To get these restrictions
 we consider  the polynomials, that enter in the expression under the square root in \eqref{preprefineq1},   and their positive
roots.

The first term under the square root is
$
Mz^{2}-1
$
and its root equals to the coordinate of the horizon in the BTZ black hole with $a=0$
\begin{equation}\label{zeroahor1}
Z_{0}=\frac{1}{\sqrt{M}}.
\end{equation}

The second term under the square root is ${\cal F}(a,M,z_m)$. It doesn't contain the integration variable $z$, but it still gives contribution to the sign of the expression under the square root and we keep it under the integral for a convenience.
The roots of  ${\cal F}(a,M,z_{m})$,
\begin{equation}\label{hor1zm}
{\cal F}(a,M,z_{m})=-1+z_m^{2}\left(-{a}^{2}z_m^{2}+M\right)=0,
\end{equation}
are given precisely by the horizons \eqref{horizon1} of the Lorentzian BTZ black hole,
\begin{equation}
z_{+}={\displaystyle \frac{1}{2}{\frac{\sqrt{2}\sqrt{M+\sqrt{{M}^{2}-4\,{a}^{2}}}}{a}}};\: z_{-}={\displaystyle \frac{1}{2}{\frac{\sqrt{2}\sqrt{M-\sqrt{{M}^{2}-4\,{a}^{2}}}}{a}}};\: Z_{0}<z_{-}<z_{+}.
\end{equation}

The third term under the square root is
$
z^{2}-z_{m}^{2}
$
and its root is
\begin{equation}\label{zzm1}
z=z_{m}.
\end{equation}

The fourth term under the square root is ${\cal G}(M,z,z_{m})$,
\begin{equation}\label{polycurve1}
{\cal G}(M,z,z_{m})=Mz_{m}^{2}z^{2}-z^{2}-z_{m}^{2}=0,
\end{equation}
and it's roots lie on the curve
\begin{equation}\label{curve1}
z=\frac{z_{m}}{\sqrt{Mz_{m}^{2}-1}}
\end{equation}
in the $(z,z_m)$ plane.

The solution of equation $\frac{z_{m}}{\sqrt{Mz_{m}^{2}-1}}=z_{m}\,$ is
$z_{m}=\sqrt{\frac{2}{M}}$ and it defines intersection of $z_m$ and curve \eqref{curve1}.

In Fig.\ref{chessboard}.A  we show the positions of  the  roots  of above equations \eqref{zeroahor1}, \eqref{hor1zm}, \eqref{zzm1} and \eqref{polycurve1} on the plane $(z,z_m)$. We divide the plane $(z,z_m)$ into  sections.
It is easy to fix the sign of the expression under the square root for small $z$ and $z_{m}$, say at the point $p$ (see Fig.\ref{chessboard}.A), where the sign is $+$.
To get the sign at any given point we connect it with the point $p$, assuming that the curve crosses the lines in normal direction. The sign is changed when the curve has crossed one of the curves drawn.

From Fig.\ref{chessboard}.A we see that $z_{m}$ can take values from $0$ to $Z_{0}$.

\subsection{ Euclidean version.}
Now we repeat all the steps of analysis made in 3.1, but in the Euclidean BTZ background.
So in this subsection we consider the expression under the square root from formula \eqref{preprefineq1} with change $a$$\rightarrow$$ia$.

The first term under the square root is
$
Mz^{2}-1
$
and its root equals to the coordinate of the horizon in the BTZ black hole with $a=0$
\begin{equation}\label{zeroahor2}
Z_{0}=\frac{1}{\sqrt{M}}.
\end{equation}

The second term under the square root is ${\cal F}(ia,M,z_m)$. It doesn't contain the integration variable $z$, but it still gives contribution to the sign of the expression under the square root and we keep it under the integral for a convenience.
The roots of  ${\cal F}(ia,M,z_{m})$,
\begin{equation}\label{hor2zm}
{\cal F}(ia,M,z_{m})=-1+z_m^{2}\left({a}^{2}z_m^{2}+M\right)=0,
\end{equation}
are precisely given by the horizons \eqref{horizon2e} of the Euclidean BTZ black hole. Note that $z^{(E)}<Z_{0}$.

The third term under the square root is
$
z^{2}-z_{m}^{2}
$
and its root is
\begin{equation}\label{zzm2}
z=z_{m}.
\end{equation}

The fourth term under the square root is ${\cal G}(M,z,z_{m})$,
\begin{equation}\label{polycurve2}
{\cal G}(M,z,z_{m})=Mz_{m}^{2}z^{2}-z^{2}-z_{m}^{2}=0,
\end{equation}
and its roots lie on the curve
\begin{equation}\label{curve2}
z=\frac{z_{m}}{\sqrt{Mz_{m}^{2}-1}}
\end{equation}
in the $(z,z_m)$ plane.

\begin{figure}[h!]
    \centering
     \includegraphics[width=7cm]{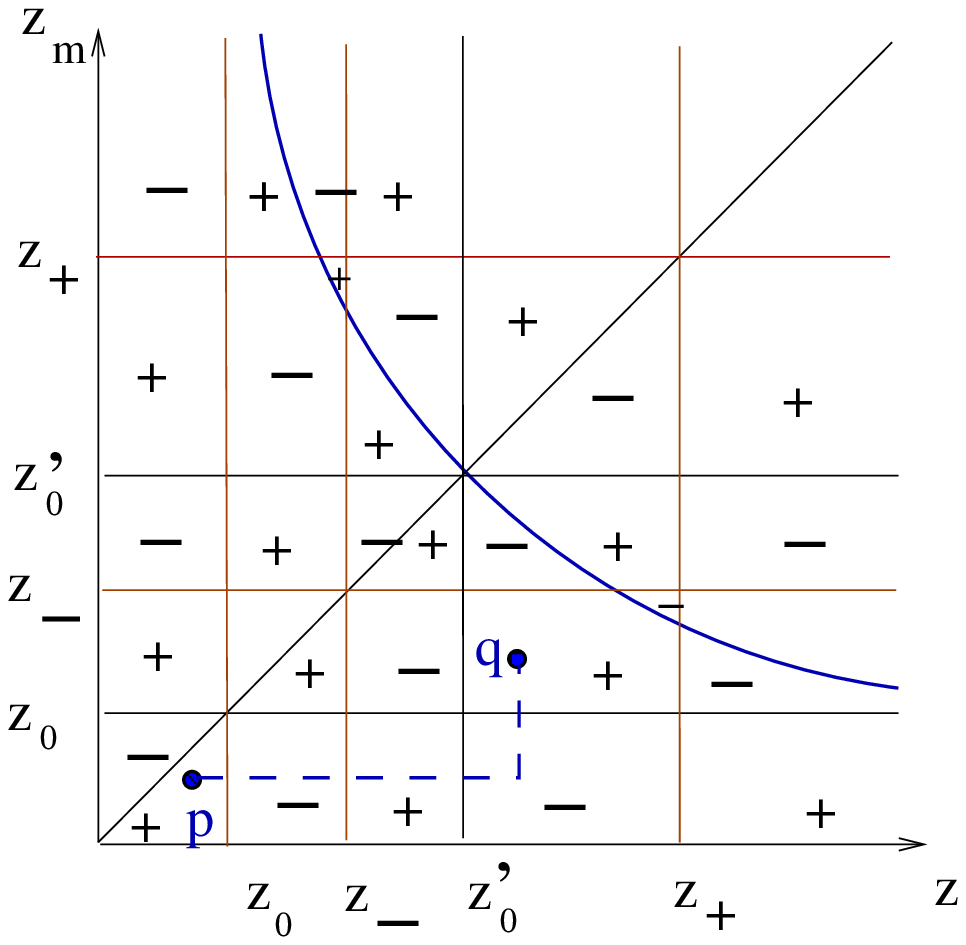}$\,\,\,\,$ $\,\,\,\,$  \includegraphics[width=7cm]{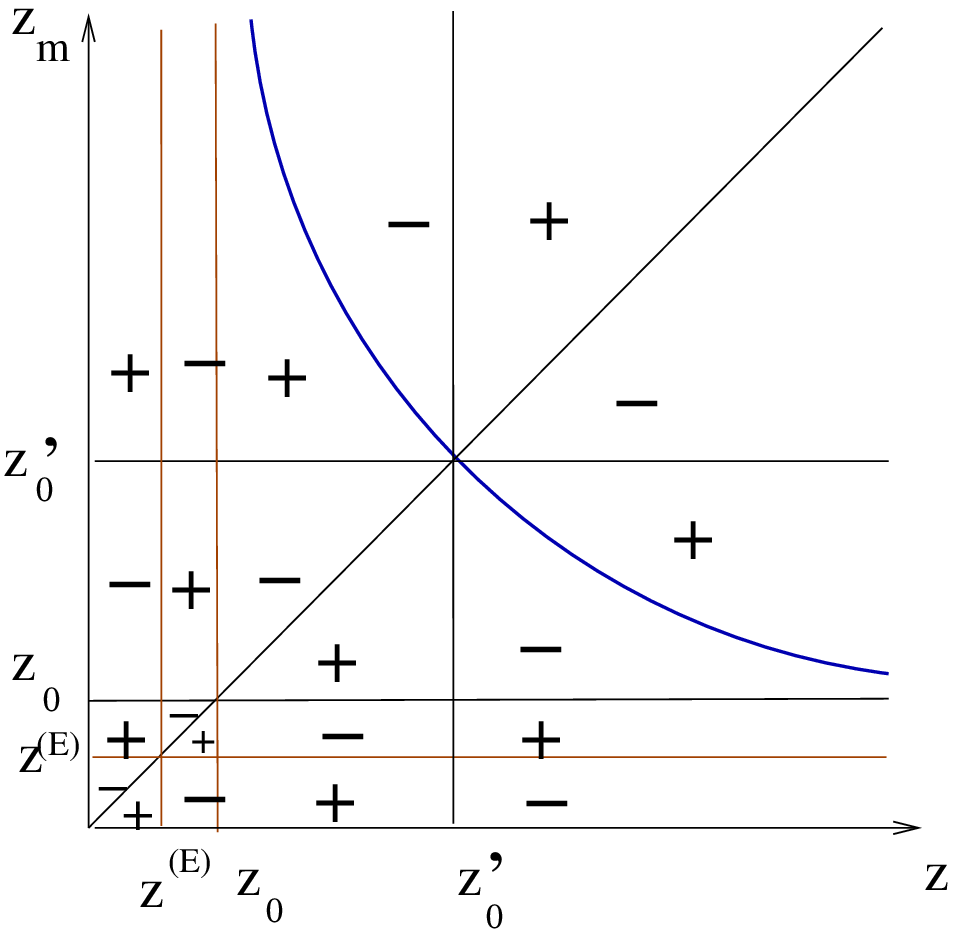}
     \caption{Roots positions. Left panel corresponds to the case of the Lorentzian BTZ, right panel corresponds to the Euclidean one. $z_0=1/\sqrt{M}$,  ${z^{\prime}}=\sqrt{2/M}$. The dashed blue curve shows a change of the sign from the point $p$ to the point $q$. The blue solid curve corresponds to curve $z=\frac{z_{m}}{\sqrt{Mz_{m}^{2}-1}}$ .}\label{chessboard}
\end{figure}

In Fig.\ref{chessboard}.B  we show the positions of  the  roots  of above equations \eqref{zeroahor2}, \eqref{hor2zm}, \eqref{zzm2} and \eqref{polycurve2} on the plane $(z,z_m)$.
From Fig.\ref{chessboard}.B we see that  in the Euclidean case  $z_{m}$ can take values only from 0 to $z^{(E)}$.

Comparing Fig.\ref{chessboard}.A and Fig.\ref{chessboard}.B  we see, that there is a crucial difference between the behaviour of the string profile in the bulks of different signature cases.

\section{Formulae for the interquark distance.}

We can approximate integral in \eqref{preprefineq1} taking into account only the main contribution from the poles at points
$z_{m}$\,and $z_{-}$, and zero at point $\frac{1}{\sqrt{M}}$ i.e. we put in\eqref{preprefineq1}  ${\cal G}(M,z_{m},z_{m})$ instead of ${\cal G}(M,z,z_{m})$ and $(-z_{+}^{2}+z_{m}^{2})$ instead $(-z_{+}^{2}+z^{2})$.

In this approximation the integral in formula \eqref{preprefineq1} takes the form

\begin{equation}\label{angle1}
\frac{\phi_{0}}{2}\approx{z_{-}z_{+}}\intop_{0}^{z_{m}}\frac{{{z}^{2}}\sqrt{(z_{-}^{2}-z_{m}^{2})(-z_{+}^{2}+z_{m}^{2})\left(M{z}^{2}-1\right)}}
{\sqrt{({z}^{2}-z_{m}^{2})({{z_{m}^{2}}}\left(M{z^2_m}-1\right)-{{z_{m}^{2}}})}(z_{-}^{2}-z^{2})(-z_{+}^{2}+z_{m}^{2})}dz.
\end{equation}

Taking the limits $z_{-}\rightarrow\frac{1}{\sqrt{M}}$,\,$z_{+}\rightarrow\infty$ in \eqref{angle1}  one can get

\begin{equation}\label{lim_angle1}
{\displaystyle {\frac{\phi_{0}}{2}\approx\intop_{0}^{z_{m}}\frac{{{z}^{2}}\sqrt{M{z_{m}^{2}}-1}}{\sqrt{(M{z}^{2}-1)({z}^{2}-{z_{m}^{2}})({{z_{m}^{2}}}\left(M{z^2_m}-1\right)-{{z_{m}^{2}}})}}}}dz.
\end{equation}

Integration of \eqref{lim_angle1} leads to the case of zero rotation formula for the interquark distance
\begin{equation}\label{lim_finangle1_M}
\frac{\phi_{0}}{2}\approx\frac{1}{Mz_{m}}\sqrt{\frac{1-M{z_{m}^{2}}}{2-M{z_{m}^{2}}}}\ {\cal {E}}({z_{m}}\sqrt{M})
\end{equation}
where ${\cal {E}}(x)$ is given in terms of the elliptic integrals
\begin{equation}
{\cal {E}}(x)={{K}\left(x\right)}-{{E}\left(x\right)},
\end{equation}
$K(x)$\, is the complete integral of the first kind and $E(x)$\, is the complete integral of the second kind \cite{stegun}.

After performing integration in the case of non-zero rotation \eqref{angle1} and some algebra we get
\begin{equation}\label{fin_angle1}
\frac{\phi_{0}}{2}\approx{z_{-}z_{+}}{\displaystyle \sqrt{\frac{\left(z_-^2-z_{m}^{2}\right)}{\left(z_+^{2}-z_m^2\right)}}\frac{\left({\cal {B}}(z_{-},z_m,M)+Mz_-^2\ {\cal {E}}({z_{m}}\sqrt{M}) \right)}{{z_m}\left(\sqrt{2-Mz_m^2}\right)}},
\end{equation}
where ${\cal {B}}(z_{-},z_m,M)$ is
\begin{equation}\label{formulaB}
{\cal {B}}(z_-,z_m,M)=\left(1-Mz_-^2\right)\Pi\left({\frac{z_m^2}{z_-^2}},{z_{m}}\sqrt{M}\right)
\end{equation}
and $\Pi(\nu,k)$ is the incomplete elliptic integral of the third kind \cite{stegun}

\begin{equation}
\Pi(\nu,k)=\intop_{0}^{1}\frac{dt}{(1-\nu t^2)\sqrt{(1-t^2)(1-k^2t^2)}}.
\end{equation}

Taking the limits $z_{-}\rightarrow\frac{1}{\sqrt{M}}$, $z_{+}\rightarrow\infty$ in \eqref{fin_angle1} one can get \eqref{lim_finangle1_M}.

For the Euclidean version of the BTZ background the formula for the interquark distance is the same up to the change of
horizons to the Euclidean one as following  ${z_-}\rightarrow {z^{(E)}}$, ${z_+}\rightarrow z^{(E)}_C$.
Note, that in this approximation for $a>\frac{M}{2}$ one should take the real part of formula \eqref{fin_angle1}.

Near $z_{m}\sim0$ the expression \eqref{fin_angle1}  is linear as it should be.
One can expand \eqref{fin_angle1} up to the higher order in $z_m$
\begin{equation}\label{app1}
\frac{\phi_{0}}{2}\approx\frac{\pi z_{m}}{2}+\frac{\sqrt{2}}{64}(2M-\sqrt{M^{2}-4a^{2}})z_{m}^{3}+O(z_{m}^{5}).
\end{equation}

For the Euclidean version of the BTZ black hole by performing a Wick rotation $a\rightarrow ia$ of expression \eqref{app1}
one gets
\begin{equation}\label{app2}
\frac{\phi_{0}}{2}\approx\frac{\pi z_{m}}{2}+\frac{\sqrt{2}}{64}(2M-\sqrt{M^{2}+4a^{2}})z_{m}^{3}+O(z_{m}^{5}).
\end{equation}

\begin{figure}[h!]
    \centering
     \includegraphics[width=9 cm]{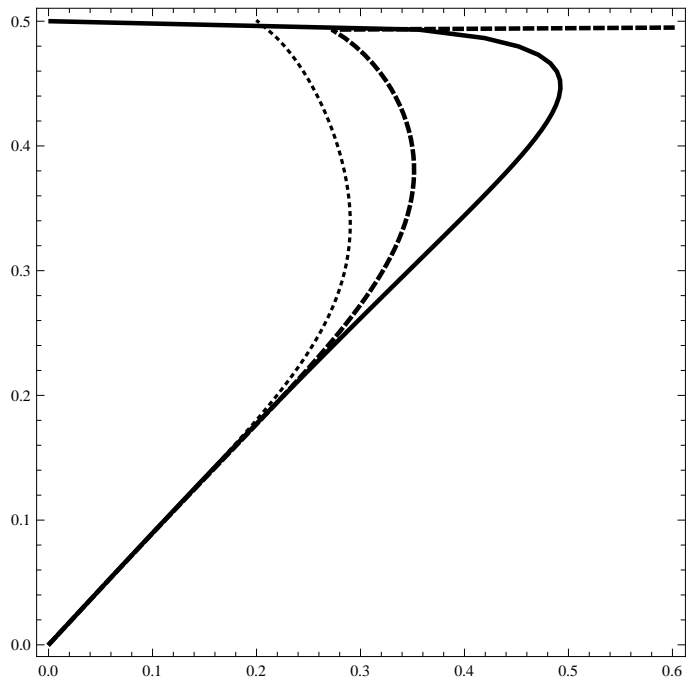}$\,\,\,\,$ $\,\,\,\,$
     \caption{The dependence of the string profile maximum on the interquark distance in the Lorentzian case of the BTZ. The solid, dashed, and
     dotted  curves correspond to $a=0.01$, $a=6$, $a=9$ respectively. Here M=4.}\label{beh_s1l}
\end{figure}

\begin{figure}[h!]
    \centering
     \includegraphics[width=9 cm]{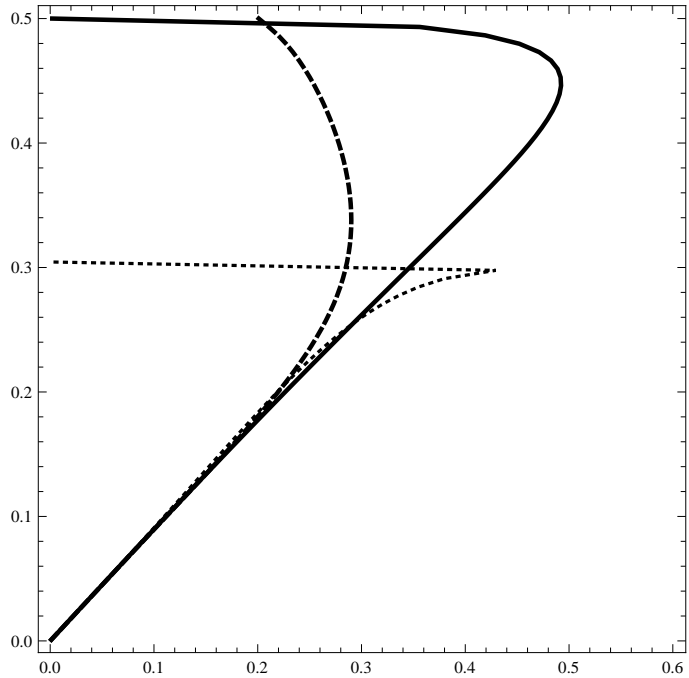}$\,\,\,\,$ $\,\,\,\,$
     \caption{ The dependence of the string profile $z_{m}$ maximum on interquark distance $\phi_{0}$ in the Lorentzian and the Euclidean versions of the BTZ black hole.
      $M=4$ is equal to 4. $a=9$ for the dashed and dotted lines, which correspond to the Lorentzian and the Euclidean case respectively. The solid line corresponds to $a=0$. }\label{beh_s2l}
\end{figure}

\begin{figure}[h!]
    \centering
     \includegraphics[width=9 cm, height=8 cm]{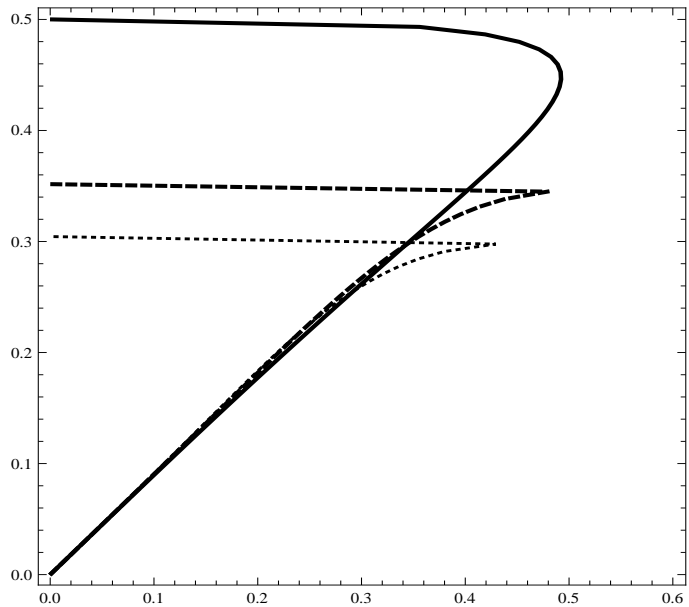}$\,\,\,\,$ $\,\,\,\,$
     \caption{The dependence of the string profile maximum on the interquark distance in the Euclidean case of the BTZ. The solid, dashed, and
     dotted curves correspond to $a=0.01$, $a=6$, $a=9$ respectively. Here M=4.}\label{beh_s3l}
\end{figure}

In Fig.\ref{beh_s1l}-\ref{beh_s3l} behaviour of the string profile in the BTZ bulk is shown for the different metric signatures and values of $a$.
The straight lines show the limits of our plots.

\section{The action and the potential.}

Using the expression for derivative $\frac{dz}{d\phi}$ one can write down the expression for the
action
\begin{equation}\label{action1}
S=-T\intop_{-\phi_{m}}^{\phi_{m}}d\phi\frac{z_{m}^{2}(-a^{2}z^{4}+Mz^{2}-1)}{z^{4}\sqrt{(a^{2}z_{m}^{4}-Mz_{m}^{2}+1)}},
\end{equation}
here $z=z(\phi)$ in accordance with string dynamics given by \eqref{P_int}.
Changing the variable of integration to $z$ (using \eqref{P_int}) one can write down the action in the following form

\begin{equation}\label{action_zm}
S=-2T\intop_{\epsilon}^{z_{m}}{\displaystyle {\frac{z_m^2\sqrt{1-M{z}^{2}}}{\sqrt{(M{z}^{2}z_m^2-{z}^{2}- z_m^2)(z^{2}-z_{m}^{2})}}}}\frac{dz}{z^{2}}.
\end{equation}
One can see, that the action doesn't contain $a$
explicitly, so the main dependence on $a$ is contained in $z_{m}$. If we take $\epsilon=0$ we get a linear divergence of the integral (\ref{action_zm}). This divergent term should be subtracted. We are going to discuss renormalizations in more details in the next section.

The potential of the interquark interaction can be obtained as
\begin{equation}\label{potential}
  V(\phi)=\frac{S_{ren}(\phi)}{T}.
\end{equation}

We are interested in the behaviour of the potential for small $z_m$.

To obtain the approximation  we expand the the integrand \eqref{action_zm} in series in $M$. Then we get
\begin{equation}\label{potential_expansion1}
V(z_m)\approx-\frac{2A}{z_m}-2CMz_m,
\end{equation}
where $A$ is constant \eqref{coeff1} and $C=\frac{1}{4}({E(i)-2K(i)})\approx0.1779$.

\begin{equation}\label{coeff1}
 A=\frac{\sqrt{2}\pi^{3/2}}{\Gamma^2(\frac{1}{4})}\approx0.599.
\end{equation}

\begin{figure}[h!]
    \centering
     \includegraphics[width=10 cm,height=9 cm]{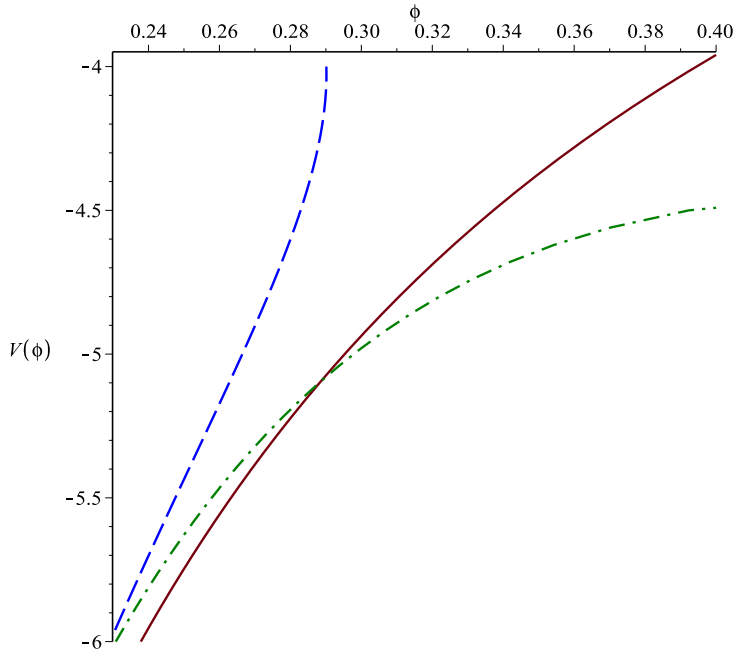}$\,\,\,\,$ $\,\,\,\,$
     \caption{ The potential for the different rotation parameter values and signatures (without finite counterterms).
      The solid red,  green dot-dashed and the blue dashed curves correspond to $a=1$, $a=9$ for the Euclidean case and $a=9$ for the Lorentzian case  respectively. Here M=4. }\label{potential_fig}
\end{figure}

Let's consider the potential of interquark interaction in the Euclidean and the Lorentzian case with $a\gg\frac{M}{2}$.
From Fig.\ref{potential_fig} one can see that the potential exhibits strong dependence on angular moment and depends on the signature of the BTZ we use.

\section{Renormalization and static  quarks in  BTZ background.}
In this section we discuss the renormalization scheme for the Lorentzian and
the Euclidean signatures. Renormalization of the string action (\ref{action_zm}
) is
related with quark self-energy and in the holographic approach is  associated
with sum of actions of two single  strings  hanging from the boundary\cite{Maldacena:1998im,Brandhuber:1998bs,Rey:1998bq}.
Namely, in the zero temperature case renormalization can be done by substraction of the static quarks action, i.e.
 $2\intop_{\epsilon}^{\infty}\frac{dz}{z^2}$, that  contains only a linear divergent part $2/\epsilon$ (compare  with the regularization used for local correlators
 \cite{Aref'eva:1998nn})
  In the  black hole background with the Euclidean signature
 one usually  subtracts the action $2\intop_{\epsilon}^{z_{H}}\frac{dz}{z^2}$, that is an action  of two strings, stretched between the boundary
  and the horizon of the black hole (here $z_{H}$ is the coordinate of the horizon).
  These strings correspond to self-energy of static quarks in thermal bath and their actions contain the  divergent part
  and finite temperature dependent part.

 In the case of the Lorentzian signature there is no well established procedure of renormalization. In \cite{Albacete:2008dz} to perform subtraction complex single string configurations
 have been used and  renormalization have been made by substraction of the
 real part  of the action of complex hanging configuration taken  at infinity.
 In this article we use a slightly  different scheme of renormalization than one used in \cite{Albacete:2008dz}.

 \subsection{The Lorentzian case.}

\begin{figure}[h!]
    \centering
     \includegraphics[width=7 cm]{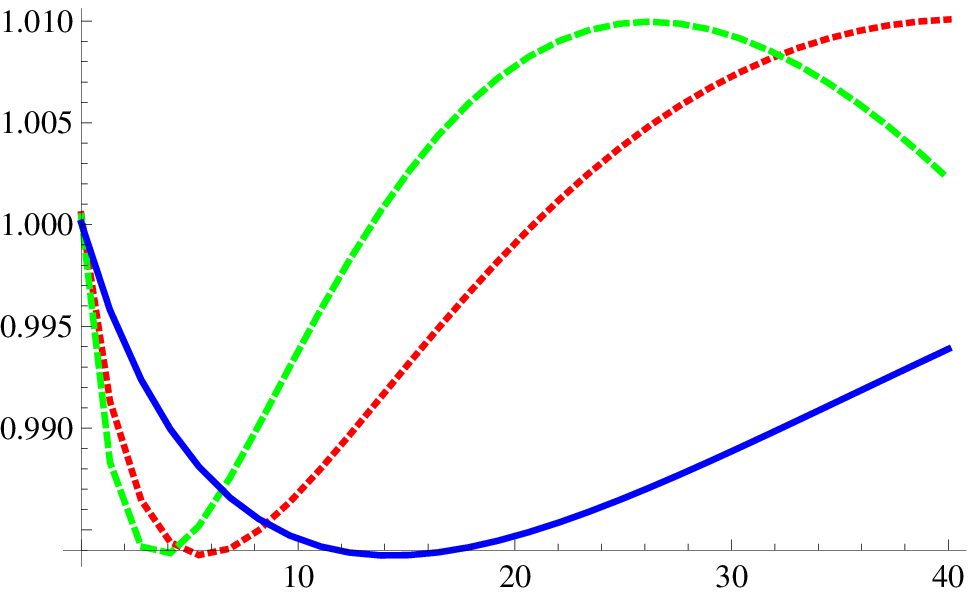}$\,\,\,\,$ $\,\,\,\,$
     \includegraphics[width=7 cm]{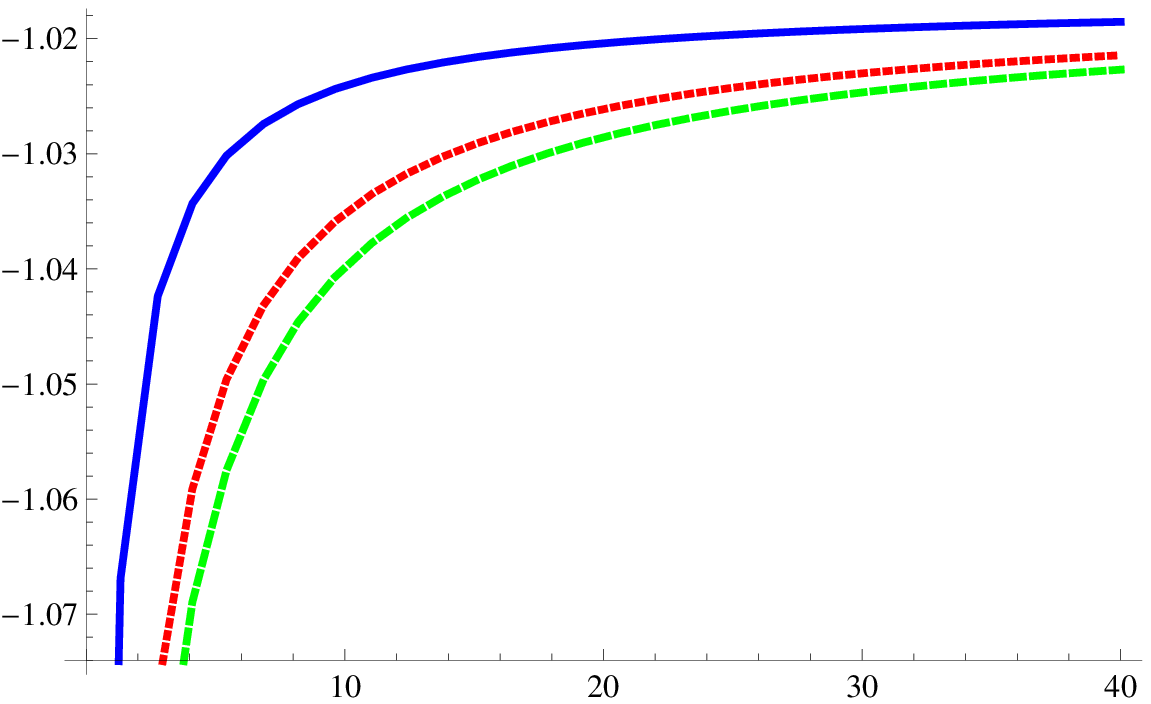}
     \caption{The dependence of the  ratios of an approximated value of $S^{(L)}_{f.p.}$  to its numerical value on rotation parameter $a$ in the Lorentzian case.
             Left panel: $M<1$. The blue solid, green dashed and the red dotted curves correspond to $M=25$, $M=16$ and $M=64$, respectively.
              Right panel: $M\ll1$, $a\gg1$. The blue solid, green dashed and the red dotted curves correspond to $M=0.64$, $M=0.81$ and $M=0.25$, respectively.}\label{estim_lor}
\end{figure}

 Let's consider the action of the single complex-valued string
 configuration at infinity.
 The real part of this action contains divergent and finite parts. For $a<\frac{M}{2}$ (i.e. below the extremal case)
   this finite part equals to zero. For the formal case $a>\frac{M}{2}$  the finite part of the single  string action $S_{f.p}$ is non zero. Indeed,
 the action of the single string that contributes in the BTZ background with mass $M$ and rotation parameter $a$ is

\begin{equation}\label{single_string_l}
S^{(L)}_{\parallel}=-2\intop_{\epsilon}^{\frac{1}{\sqrt{M}}}\sqrt{\frac{1-Mz^2}{1-Mz^2+{a^2}z^4}}
\,\frac{ dz}{z^2}=-(\frac{2}{\epsilon}+2S^{(L)}_{f.p.})
\end{equation}
We numerically estimate $S^{(L)}_{f.p.}$ for $M\gg1$  as $S^{(L)}_{f.p.}\approx -\sqrt{M}- a\tan{\left(\frac{8}{3\sqrt{M}}\arctan{\frac{1}{5.4}}\right)}$.
For the values $M<1$ and $a>1$ one can estimate the finite part of the action as $S^{(L)}_{f.p.}\approx  -\frac{1.2}{z_{h}}$.
In Fig.\ref{estim_lor} the comparison of the approximated value of $S^{(L)}_{f.p.}$  and its numerical value  is presented.

This scheme of renormalization has a natural relation to the method of the complex-valued
 strings and is similar to the method we consider in the case of the Euclidean signature.
  In the zero-temperature limit results obtained from this method coincide with the standard results.
\subsection{The Euclidean case.}

The substraction scheme in the Euclidean case is standard, so the action of string that takes part in renormalization is
\begin{equation}\label{single_string_e}
S^{(E)}_{\parallel}=2\intop_{\epsilon}^{z_{E}}
\sqrt{\frac{1-Mz^2}{1-Mz^2-{a^2}z^4}}\,\frac{dz}{z^2}=\frac{2}{\epsilon}+2S^{(E)}_{f.p},
\end{equation}
where $z_{E}$ is the Euclidean horizon, $S^{(E)}_{f.p}$ is the finite part of the action
$S_{\parallel}$.

For the values $M<1$ we get numerically the finite part  $S^{(E)}_{f.p.}\approx -\frac{1.5}{z_{h}}$.
For the values $M\gg1$  we get numerically that $S^{(E)}_{f.p.}\approx -\sqrt{M}$.

\begin{figure}[h!]
    \centering
     \includegraphics[width=7 cm]{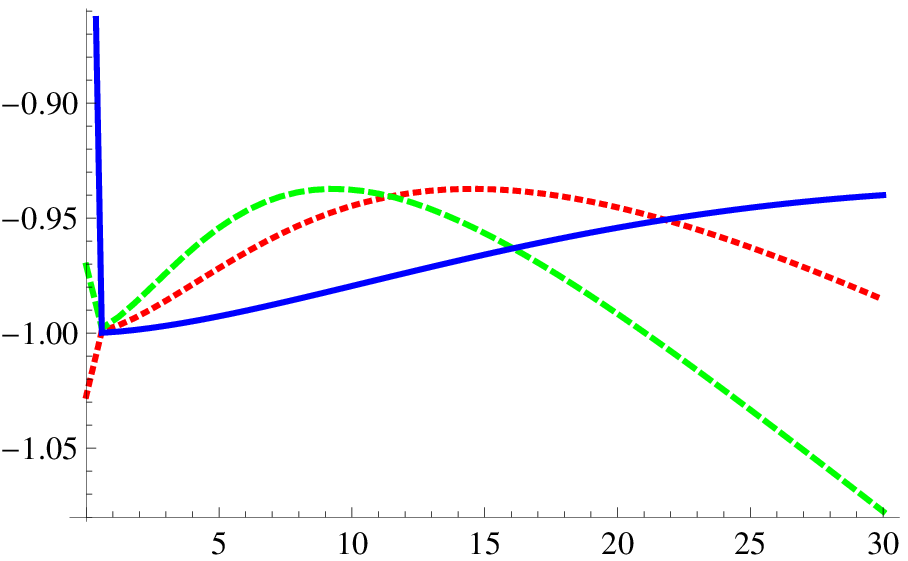}$\,\,\,\,$ $\,\,\,\,$ \includegraphics[width=7 cm]{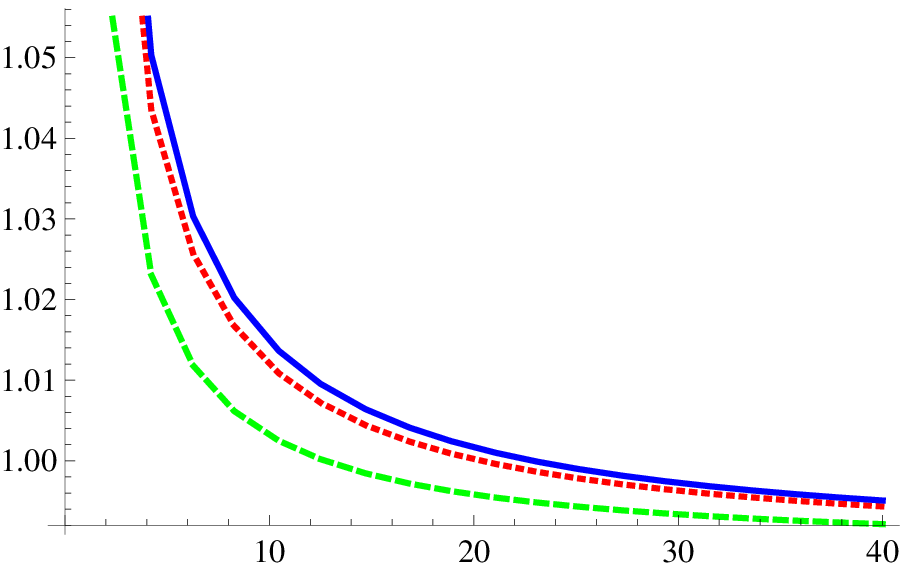}
     \caption{The dependence of the  ratios of the estimation and the numerical results on rotation parameter $a$ for the $S^{(E)}_{f.p}$ in the Euclidean case.
             Left panel: $M\gg1$. The blue solid, green dashed and the red dotted curves correspond to $M=25$, $M=16$ and $M=64$, respectively. Right panel: $M<1$. The blue solid, green dashed and the red dotted curves correspond to $M=0.8$, $M=0.5$ and $M=0.9$, respectively. }\label{estim_euc}
\end{figure}
\newpage
\section{Conclusions}
In this work  using the dual prescription the singlet potential of the quark-antiquark pair in QGP with non-zero angular momentum has been calculated.
 To get this we have considered non-local operators on boundary of the BTZ black hole with non-zero rotation parameter.
 We have got different results for singlet potentials in a different signatures of the BTZ.
We relating the results obtained in the Lorentzian signature with real-time thermal correlators. For the Euclidean signature the results should be related with the Euclidean thermal correlators. Therefore, we get that potentials extracted,
via chosen holographic prescription, from Wilson and Polyakov loops are different in the case of QGP with non-zero angular momentum.

\section*{Acknowledgments}
This work was partially supported by the Russian Foundation for Basic Research grant 14-01-00707-a  and Russian Federation President grant for support
of young scientists, grant MK-2510.2014.1 (D.A)

\appendix
\section{ Comparison of the numerical calculation and the analytic\\ formulae.}
In this appendix we compare the analytical expressions for the string profile behaviour and potential with numerical calculation in
the different signatures, rotation parameter $a$ and mass $M$ values.

\begin{figure}[h!]
    \centering
     \includegraphics[width=7cm]{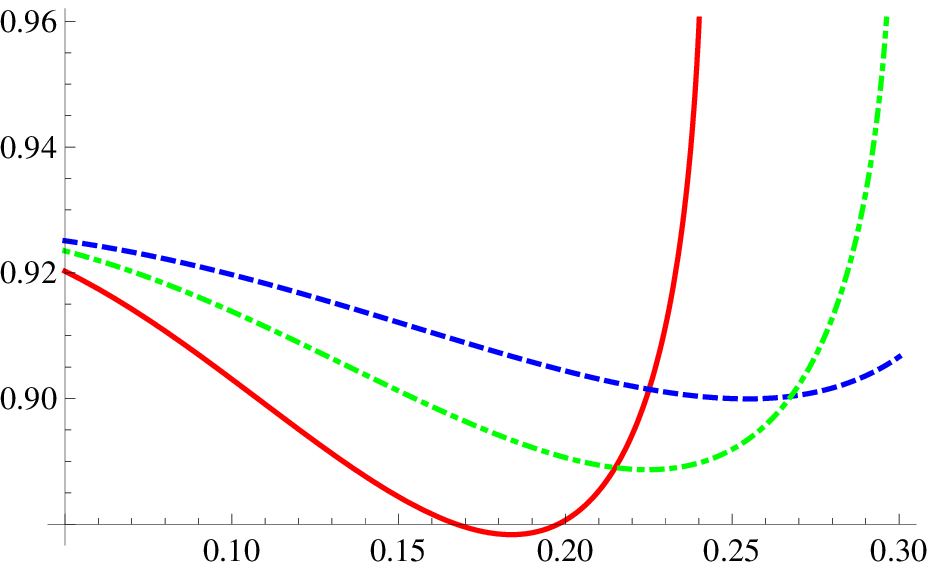}$\,\,\,\,$ $\,\,\,\,$ \includegraphics[width=7cm]{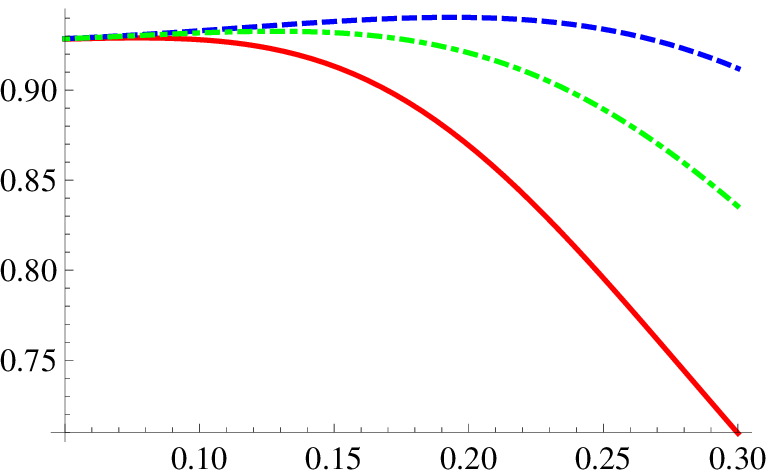}
     \caption{The dependence of the  ratios of the numerical and the approximated values of  the interquark distance $\phi$ on $z_m$, the string profile maximum coordinate.
              The blue dashed, green dot-dashed and the red solid curves correspond to $a=6$, $a=9$ and $a=15$, respectively. Here M=4. Left panel: the Euclidean signature. Right panel: the Lorentzian signature.}\label{num2}
\end{figure}

 Fig.\ref{num2} demonstrates a comparison of  analytical formula \eqref{fin_angle1} and numerically evaluated integral \eqref{preprefineq1} for the different rotation parameter values for the different signatures. Note, that in the Lorentzian signature the analytical approximation is worse than in the Euclidean one.

\begin{figure}[h!]
    \centering
    \includegraphics[width=9 cm]{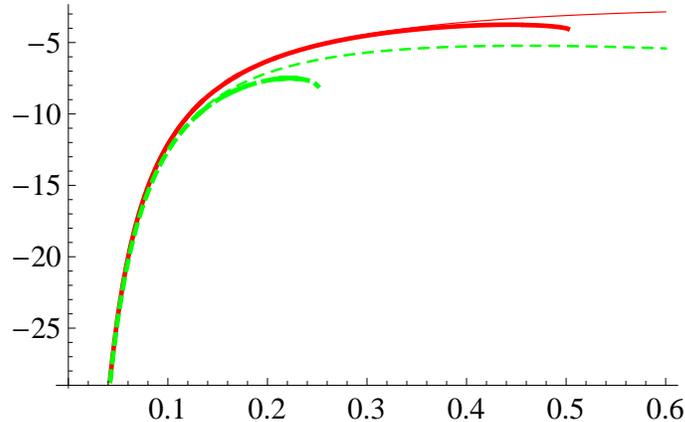}
    \caption{The dependence of the potential $V(z_m)$  on the string profile maximum $z_m$ coordinate (without the finite counterparts addition).
    The thick lines correspond to the numerical calculation, the thin ones correspond to analytical formula.
     The red and green lines correspond to $M=4$ and $M=16$ respectively.}\label{num3}
\end{figure}

Fig.\ref{num3} demonstrates a comparison of  formulae  \eqref{potential_expansion1} and
 numerically evaluated integral \eqref{action_zm} for the different rotation parameter values.

\end{document}